\documentclass[journal]{IEEEtran}

\ifCLASSINFOpdf
\else
   \usepackage[dvips]{graphicx}
\fi
\usepackage{url}

\usepackage{graphicx}
\usepackage{amsmath}
\usepackage{xcolor}
\usepackage{subfigure}

\renewcommand{\arraystretch}{1.5}

\begin{document}



\title{Poisson Informed Retinex Network for Extreme \\ Low-Light Image Enhancement}

\author{Isha Rao, Ratul Chakraborty, and Sanjay Ghosh, \IEEEmembership{Senior Member, IEEE}
\thanks{This work is supported by the Faculty Start-up Research Grant (FSRG), Indian Institute of Technology Kharagpur, India.}
\thanks{Isha Rao, Ratul Chakraborty, and Sanjay Ghosh are with Department of Electrical Engineering, Indian Institute of Technology Kharagpur, WB 721302, India (e-mail: \texttt{sanjay.ghosh@ee.iitkgp.ac.in}).}}

\markboth{under Review}
{Rao \MakeLowercase{\textit{et al.}}: Bare Demo of IEEEtran.cls for IEEE Journals}

\maketitle

\begin{abstract}
Low-light image denoising and enhancement is a critical task in computer vision, with applications in photography, surveillance, autonomous driving, and medical imaging. 
In such conditions, images often suffer from severe degradation due to low photon counts, making both enhancement and denoising particularly challenging. 
In many real-world  scenes of low-light imaging, noise, which is in majority 
signal-dependent, is better represented as Poisson noise. 
In this work, we address the problem of denoising images degraded by Poisson noise under extreme low-light conditions. 
%
We propose a novel Poisson noise modeling for extreme low-light images, which is then directly integrated into a trainable deep learning network for low-light image enhancement. We propose an end-to-end deep learning method based on Poisson noise informed retinex decomposition.
%
Our proposed method simultaneously enhances illumination and suppresses noise by incorporating a Poisson denoising loss that effectively addresses signal-dependent noise. Without any prior requirement for reflectance and illumination, the network learns an effective decomposition process while ensuring consistent reflectance and smooth illumination without introducing any form of color distortion. The experimental results demonstrate the effectiveness and practicality of the proposed low-light illumination enhancement method.
Our method significantly improves visibility and brightness in low-light conditions, while preserving image structure and color constancy under ambient illumination.

\end{abstract}

\begin{IEEEkeywords}
Low-light imaging, image denoising, Poisson noise, retinex model, color distortion. 
\end{IEEEkeywords}

\IEEEpeerreviewmaketitle

\section{Introduction}

\IEEEPARstart{I}{mages} captured under low-light conditions often suffer from high noise levels, poor contrast, and distorted illumination,
which result in significant degradation in the visual quality and make it difficult to extract meaningful information from the images. This issue is prominently affecting real-time systems, such as surveillance, autonomous driving \cite{grigorescu2020survey}, object detection \cite{shrikhande2023face} where accurate and clear image representation is essential for decision making. Therefore, efficient computational techniques are particularly important for reducing the loss of system performance caused by low-light images.

In recent decades, numerous conventional approaches for low-light image enhancement have been developed. Among them, traditional enhancement techniques such as histogram equalization \cite{chen2003minimum, abdullah2007dynamic, sen2010automatic} which enhances image contrast and brightness by redistributing the intensity levels of image pixels.
These histogram equalization methods often have notable limitations, such as over-enhancement and noise amplification, particularly in extremely low-light conditions \cite{ibrahim2007brightness, ma2023low}.
Based on the Retinex model \cite{land1977retinex},  researchers introduced single-scale Retinex  \cite{jobson1997properties} and multi-scale  Retinex \cite{jobson1997multiscale}  for the low-light image enhancement.
The core idea of the Retinex-based model is to decompose an image into reflectance and illumination components and then recover the illumination while preserving the reflectance, improving brightness and contrast \cite{elad2005retinex, ghosh2019fast}. 
The authors in \cite{guo2016lime} proposed to enhance image illumination by estimating a structured illumination map \cite{guo2016lime}.
Among the deep learning-based methods \cite{zhang2023learning,shi2022unsupervised,xu2022structure,huang2022towards,li2023pixel,wang2023enlighten, wang2023multi,xie2024residual, wang2024extracting,he2025zero}, RetinexNet \cite{Chen2018Retinex}, which enhances image contrast through brightness maps estimation and adjustment, with subsequent denoising using block matching and 3D filtering \cite{danielyan2011bm3d}. 
Ko et al. proposed a lightweight deep neural network model through knowledge distillation \cite{ko2021learning}. The authors in \cite{yang2022rethinking,xu2022rawformer} proposed  vision transformer-based generative adversarial network to enhance low-light images.


Conventional image enhancement techniques based on Retinex decomposition such as \cite{Chen2018Retinex}, \cite{guo2016lime,shen2022blind}, often assume that noise is independent of the signal and is modeled as zero-mean additive Gaussian noise \cite{Guan2019NODEEL}. Hence all conventional methods undergoes enhancement and then denoising of images, handling both the tasks as individual and independent of each other. However, in practical scenarios such as low-light imaging the noise is typically signal-dependent and hence is better represented by Poisson statistics \cite{hasinoff2021photon,gnanasambandam2024secrets}. 
The noise in extreme low-light images is dominated by Poisson statistics, which differs from the Gaussian noise assumption made by many standard denoising algorithms \cite{foi2008practical, niknejad2018poisson}.
This distinction makes conventional methods less effective in handling the noise typically found in low-light environments \cite{wang2017fast}. The illumination in such images is often distorted due to extreme low illumination, making it challenging to recover true scene details. 
As a result, the performance of most existing  techniques is limited under low-light conditions, where the signal-to-noise ratio (SNR) is low and photon counts are limited \cite{talbot2009efficient,kong2021low, huang2022quaternion}. 



In this paper, we propose a Poisson-aware loss function for illumination recovery and enhancement of images corrupted by Poisson noise due to poor lighting conditions. In particular,
we introduce a unified light-weight encoder-decoder network which integrates Retinex based decomposition with Poisson denoising.  Our encoder-decoder network learns an effective decomposition process while ensuring consistent reflectance and smooth illumination components. Our proposed modified Retinex decomposition has an additional component representing Poisson noise at each pixel. Our novelty is that we have a dedicated loss function in our deep learning architecture to suppress the noise component in the output/enhanced image. 


\begin{enumerate}
    \item We propose a Poisson noise modeling, which is then directly integrated into a trainable deep learning network for low-light image enhancement. By strengthening Retinex decomposition with a physics driven solution, we effectively address the challenges of insufficient information and severe feature degradation in single-image low-light conditions.

    \item By incorporating a dedicated noise estimation branch and formulating a noise-aware loss, we enable joint learning of enhancement and denoising in an end-to-end manner. Our composite loss function jointly performs enhancement and denoising of low-light images.

    \item We extensively experimented on the recovery of images under extreme low-light conditions of varied severity of lack of illumination. Handling the complexities of signal-dependent noise while preserving structure and color consistency, we are able to achieve superior performance under extreme low-light conditions.
 
\end{enumerate}

 The rest of the paper is organized as follows. In Section \ref{sec:Poisson}, we introduce the Poisson denoising problem and our proposed Poisson guided decomposition of extreme low-light image. We present our proposed deep learning network in Section \ref{sec:method}
 and the experimental results in Section \ref{sec:result}.
 Finally, we  conclude with a summary in Section \ref{sec:conc}.

\section{Related Works}

\subsection{Classical Methods}

In low-light conditions, image quality is often constrained
by the optical signal attenuation, resulting in a significant degradation of the signal-to-noise ratio and a notable increase in noise. Early enhancement methods such as histogram equalization and gamma correction aimed to stretch the dynamic range of low-light images. These methods are fast but often cause artifacts like over-enhancement and loss of detail. Adaptive versions like brightness preserving dynamic histogram equalization  \cite{chen2003minimum, abdullah2007dynamic, sen2010automatic} tried to address these issues but lacked robustness in extreme darkness.
Inspired by the human visual system, Retinex theory \cite{land1977retinex} models an image as the product of reflectance and illumination. Early works like SSR and MSR \cite{jobson1997multiscale}, \cite{jobson1997properties} used handcrafted filters, while later methods integrated bilateral filtering \cite{elad2005retinex} or bright-pass filtering \cite{ghosh2019fast}. However, these models typically assume ideal lighting and do not address noise explicitly. Based on Retinex Theory, LIME \cite{guo2016lime} enhances low-light images by estimating a structure-aware illumination map and applying pixel-wise amplification based on the estimated illumination. While this strategy improves visibility, the method does not incorporate an explicit denoising component within its core algorithm. As a result, it often amplifies the noise inherently present in low-light images. Although the authors recommend using an external denoiser such as BM3D \cite{danielyan2011bm3d} as a post-processing step, this decoupled approach lacks adaptability to varying noise levels and image structures. 

\subsection{Deep Learning Methods}
Deep learning-based methods \cite{zhao2025deep,liu2024low,zhao2024non,fei2023generative,liang2022self} have significantly advanced low-light image enhancement by learning to restore visibility and contrast directly from data.
Deep models such as RetinexNet \cite{Chen2018Retinex} use learned decomposition followed by denoising filters. However, the denoising step is handled externally using BM3D \cite{danielyan2011bm3d}, which assumes additive Gaussian noise and is decoupled from the learning process, making it suboptimal for signal-dependent noise encountered in real-world low-light scenarios.
Zero-DCE \cite{guo2020zero} proposed a zero-reference framework that learns pixel-wise brightness adjustment curves without paired supervision. While efficient, it lacks explicit modeling of structural information or noise, often resulting in overexposed or color-distorted outputs under extremely poor illumination. Similarly, EnlightenGAN \cite{9334429} uses adversarial learning to produce visually appealing results, but it tends to hallucinate textures and is prone to generating inconsistent enhancements.

Recent approaches such as Transformer-GAN \cite{yang2022rethinking} and RawFormer \cite{xu2022rawformer} employ vision transformers to capture long-range dependencies, improving structure preservation and global enhancement. Despite their strong performance, these models are computationally expensive and not well-suited for real-time or low-power applications. On the other hand, lightweight models \cite{ko2021learning, Zhang2023SingleLayer} aim to reduce complexity using shallow networks or knowledge distillation. However, they often sacrifice enhancement quality and provide limited noise suppression, particularly under extreme low-light where signal-to-noise ratios are very low.
While many existing low-light enhancement methods operate under the assumption of additive white Gaussian noise (AWGN), real-world low-light images are primarily degraded by Poisson noise, which is signal-dependent in nature. This is particularly prominent in scenarios with low photon counts, where each pixel’s intensity corresponds to a discrete number of photon arrivals.

\subsection{Methods for Extreme Low-light Images}

To better address this, several recent methods have begun incorporating Poisson noise models into their frameworks. For instance, Kong et al.\cite{Kong2021PoissonRetinex} proposed a Poisson-aware Retinex model that integrates noise-aware constraints into the decomposition of an image into illumination and reflectance. Their method formulates a joint optimization problem that regularizes illumination estimation while accounting for signal-dependent noise. While this approach improves performance under moderate low-light settings, it relies on handcrafted priors and solves the problem through iterative optimization. This makes it computationally inefficient and less suitable for real-time or end-to-end training.

Similarly, Huang et al. \cite{Huang2022QuaternionPoisson} introduced a quaternion-screened Poisson equation framework for low-light image enhancement. They extended the screened Poisson equation to quaternion space to jointly process color channels, allowing more consistent illumination correction and smoother gradient preservation. Although their method demonstrates improved color fidelity and brightness restoration, it does not perform explicit denoising and is primarily focused on the enhancement of structural components. Additionally, like other model-based approaches, it lacks learning adaptability and does not support data-driven optimization or joint image reconstruction.


\section{Poisson denoising problem and Beyond}
\label{sec:Poisson}
The primary source of noise in digital images arises during image acquisition or transmission. In the case of low-light images with poor exposure, noise is predominantly due to a low signal-to-noise ratio (SNR) and low photon count—fewer photons reaching the sensor limit the number of photons collected by a pixel. This results in increased uncertainty, as described by the Poisson distribution, which models the occurrence of independent random events such as photon arrivals. Consequently, Poisson noise, also known as shot noise, becomes the dominant form of noise in such conditions.

Let $Y$ denotes a noisy image produced by a sensor. The objective of denoising is to recover a ground-truth clean image $X$ observed by the sensor. In extreme low-light imaging, the noise is dominated by Poisson noise. Given the true value $X(i)$ of the $i$-th pixel expressed in the number of photons, the corresponding value of the observed pixel $Y(i)$ is an independent Poisson distributed random variable;
%
\begin{equation}
    P(Y(i) = n | X(i) = \lambda) = \frac{\lambda^{n}e^{-\lambda}}{n!},
    \label{eq:Poisson}
\end{equation}
where $\lambda >0 $ is both mean and variance \cite{hines2008probability}.  
In defining the Poisson statistics for each pixel of the low light image, it is assumed that the number of arrivals during non-overlapping time intervals are independent random variables. There exists a positive quantity which is the contribution of a photon to the intensity of the pixel in \( \Delta t\) duration and satisfying the postulates for the distribution of Poisson for each value of the pixels \cite{hines2008probability}.
%

\begin{figure*}
            \centering
          \includegraphics[width=7.2in]{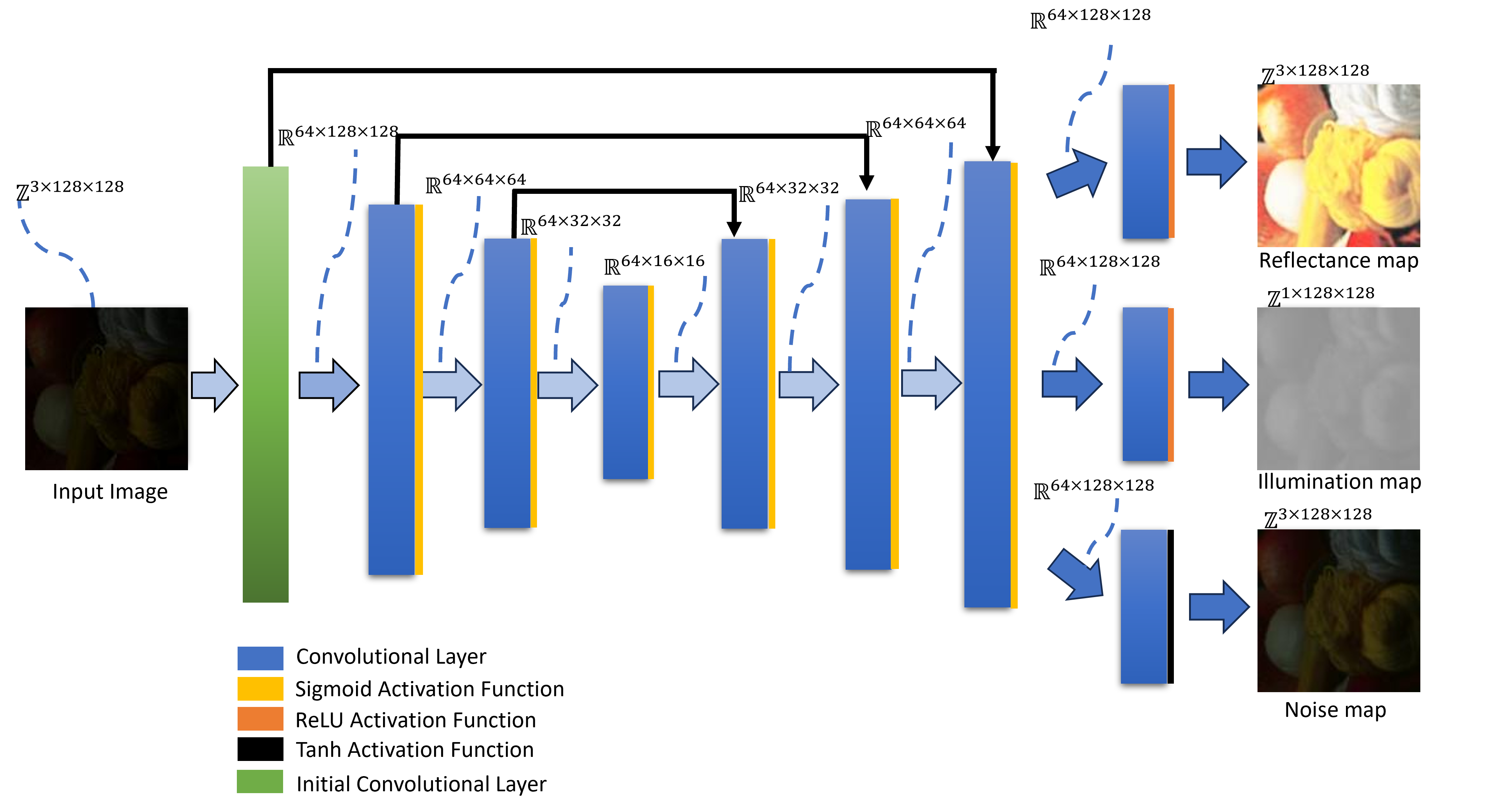}
            \caption{Proposed architecture for Poisson noise removal and low-light image enhancement. The network is structured with a convolutional encoder-decoder framework. The three branches of the output layers generate an illumination map, a reflectance map, and a noise map.}
        \label{fig:denoiseNet}
\end{figure*}

\subsection{Physics of Poisson Noise via Multiplicative Decomposition}

It is well known that low-light digital images captured by cameras suffer from high levels of Poisson noise. These images are typically captured and stored in RGB format \cite{lukac2005color}, and since Poisson noise is signal-dependent, the random nature of photon arrivals at the sensor affects all three color channels. The color image we see is a linear combination of the RGB channels obtained via the Bayer pattern, which can lead to color shifts and failure to preserve color constancy \cite{sonka2013image}, \cite{ma2024low} during image enhancement if denoising is not properly addressed \cite{parabextreme}. Performing denoising after enhancement is also ineffective in such cases, as it can further degrade the image. Therefore, to preserve color constancy under ambient illumination, it is crucial to apply denoising alongside enhancement.

Existing methods for Poisson denoising often employ a Bayesian framework, treating it as an ill-posed problem to estimate both the illumination and reflectance components from low-light images. Typically, these methods adopt the maximum a posteriori probability (MAP) approach to estimate a noise-free image \cite{niknejad2018poisson}, \cite{8588010}. Poisson noise is neither purely additive nor multiplicative, but it is signal-dependent. In extreme low-light situations, Poisson noise can become so dominant that the entire observed pixel can be considered Poisson-corrupted, and it can be formulated as a multiplicative component of the noise in the pixel values.

To address this and preserve both color constancy and prevent color shifts, we have formulated a Poisson-aware Retinex model for processing extreme low-light images. 
As it is independent random process, so individual Poisson corrupted pixel is represented as
\begin{equation}
    Y(i) = Poisson(I_{i}), \forall i,  
\end{equation}
where $i$ is the index of pixels in the image. 
From \eqref{eq:Poisson} we get
\begin{equation}
    Y(i) = \frac{(X(i))^n e^{- X(i) }}{n!}, \forall i. 
\end{equation}
Then, we expand as follows:
\begin{equation}
    Y(i) = X(i) \frac{(X(i))^{(n-1)} e^{- X(i) }}{n!}, \forall i. 
\end{equation}
Thus, the corrupted low-light image is now expressed as product of the noise-fee image $X$ and a component of Poisson noise expressed as:
\begin{equation}
    N (i) =  \frac{(X(i))^{n-1}e^{- X(i) }}{n!}, \forall i. 
\end{equation}
Finally, the observed image at the $i$-th pixel is given by
\begin{equation}
    Y(i) = X(i) N (i), \forall i.
\end{equation}
So, the low-exposure image $Y$ is represented as element-wise multiplication of the sufficiently exposed image $X$ and the multiplicative component of Poisson noise $N$. 

Recalling the retinex theory, an image can be decomposed into element-wise multiplication of illuminance \(L\) and reflectance \(R\). Therefore,
\begin{equation}
    X = L\circ R
\end{equation}
and
\begin{equation}
    Y = L \circ R \circ N.
\end{equation}
Finally, by using our hypothesis of Poisson-aware retinex method, a low light image can be decomposed into element wise multiplication of three componets that are multiplicative components of  noise \(N\), reflectance \(R\) and illuminance \(L\).  Here $\circ$ refers element-wise multiplication of two matrices.

\subsection{Poisson Noise Guided Color Image Decomposition}


It is well known that low-light digital images are typically captured and stored in RGB format \cite{lukac2005color}, and since Poisson noise at the sensor affects all three color channels. This can lead to color shifts and failure to preserve color constancy \cite{sonka2013image}, \cite{ma2024low} during image enhancement.
To preserve color constancy under ambient illumination, we consider to decompose the captured color image $X$ as following:
\begin{equation}
    Y^{c}(i) = L(i) \circ R^c (i) \circ N^c(i), \quad c \in \{r, g, b\},
\end{equation}
where $\circ$ indicates element-wise multiplication at each pixel $i$; $L$, $R$, and $N$ are illumination, reflectance, and noise components respectively.
%
Poisson noise is neither additive nor multiplicative, but it is signal-dependent. However, it can be formulated as a multiplicative component of the noise in the pixel values.
To preserve color constancy and prevent color shifts during enhancement, we formulate a Poisson guided image decomposition framework where an image is decomposed into the element-wise multiplication of three components: illumination \(L\), reflectance ($R$), and noise ($N$).

\section{Proposed Method}
\label{sec:method}

\subsection{Network Architecture}

We propose a convolutional encoder-decoder architecture that is shown in Figure \ref{fig:denoiseNet}. The key components are as follows:

\subsubsection{Initial Convolution Layer}
The first layer is a convolutional layer with 3 input channels (RGB input image patches) of size \(3\times128\times128\) and \(64\) output channels, utilizing a \(3\times3\) kernel size. This layer extracts basic low-level features from the input image, such as edges and textures. 

\subsubsection{Encoder-Decoder Layers}
The encoder consists of three consecutive convolutional layers with strides of two. 
We extract hierarchical features at multiple scales by progressively downscaling the features in each cascaded convolutional layer.
Each such layer is followed by a ReLU activation function to introduce nonlinearity for better feature representation. 

In the decoder stage, we perform feature concatenation via  skip connections. This allows for the propagation of fine-grained features from the encoder to the decoder. The feature concatenation operation results in better preservation of details during reconstruction.
The decoder consists of three deconvolutional layers. Each of these layers takes input from both the corresponding encoder layer (via concatenation) and the previous decoder layer. 
This enables the network to recover spatial resolution while combining high-level and low-level features for better image reconstruction.


\begin{figure*}
            \centering
          \includegraphics[width=0.65\linewidth]{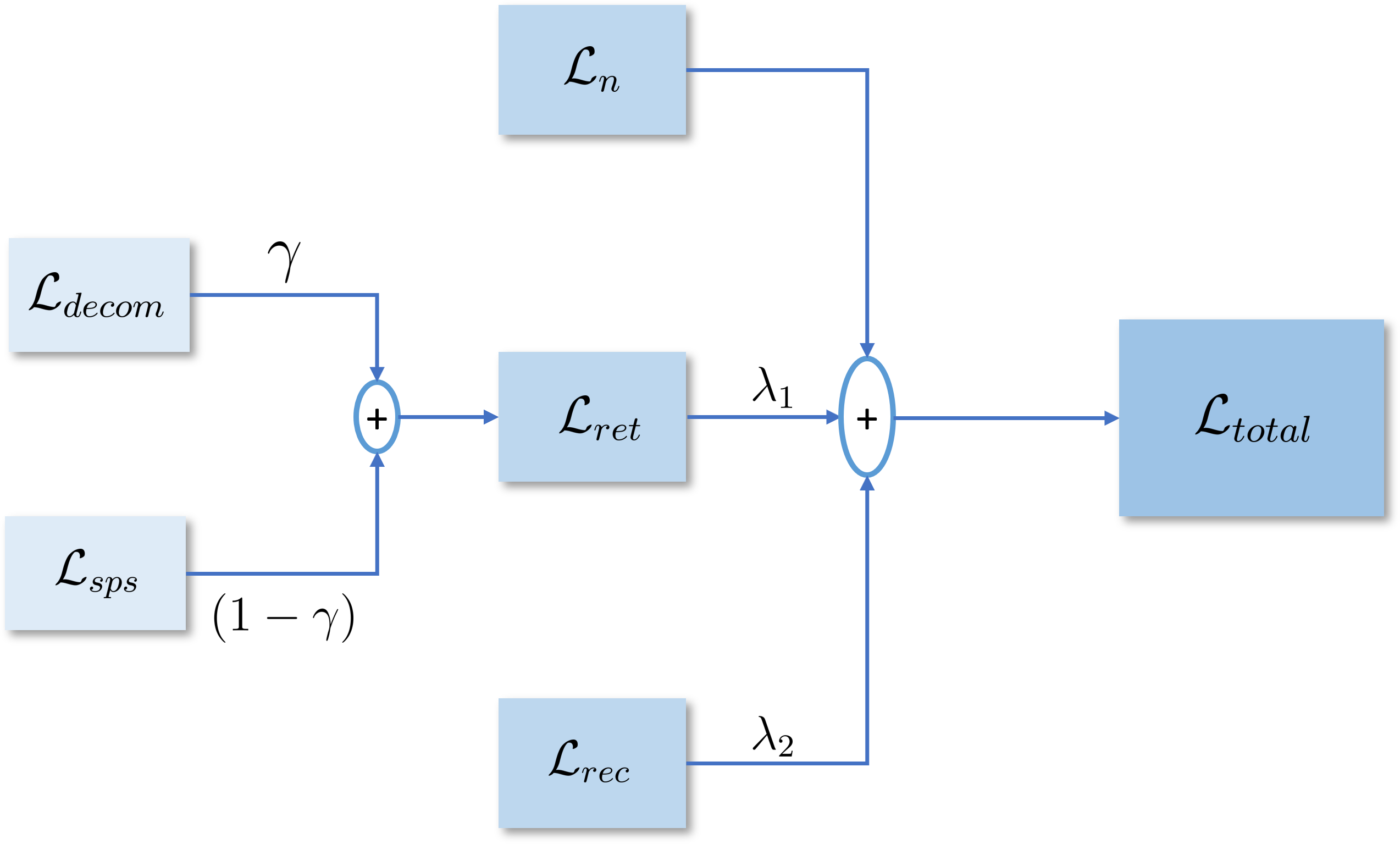}
            \caption{Schematic of the loss function in our proposed method \textbf{P}oisson \textbf{I}nformed \textbf{R}etinex of \textbf{L}ow-light images (PIRL).}
        \label{fig:loss}
\end{figure*}


\begin{figure*}
    \centering
    \setlength{\tabcolsep}{2pt}
    \renewcommand{\arraystretch}{1.2}
    \begin{tabular}{ccccc}
       (a) \textbf{Input} & (b) \textbf{LIME} & (c) \textbf{Deep-Retinex} & (d) \textbf{Zero-DCE} & (e) \textbf{Proposed} \\
        \includegraphics[width=0.19\linewidth]{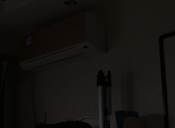} & 
        \includegraphics[width=0.19\linewidth]{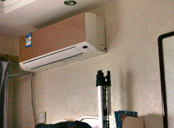} & 
        \includegraphics[width=0.19\linewidth]{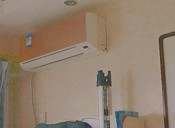} & 
        \includegraphics[width=0.19\linewidth]{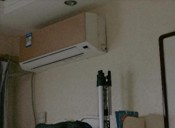} & 
        \includegraphics[width=0.19\linewidth]{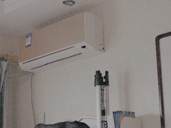} \\
        \texttt{Low-light level} $1$. \\
        \includegraphics[width=0.19\linewidth]{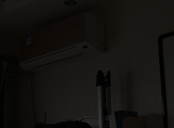} & 
        \includegraphics[width=0.19\linewidth]{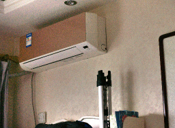} & 
        \includegraphics[width=0.19\linewidth]{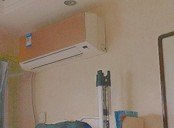} & 
        \includegraphics[width=0.19\linewidth]{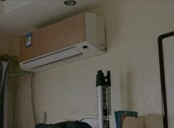} & 
        \includegraphics[width=0.19\linewidth]{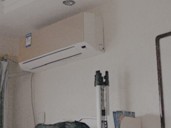} \\
        \texttt{Low-light level} $2$. \\
        \includegraphics[width=0.19\linewidth]{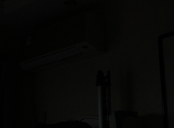} & 
        \includegraphics[width=0.19\linewidth]{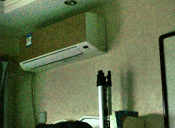} & 
        \includegraphics[width=0.19\linewidth]{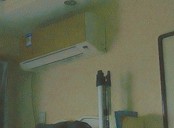} & 
        \includegraphics[width=0.19\linewidth]{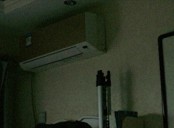} & 
        \includegraphics[width=0.19\linewidth]{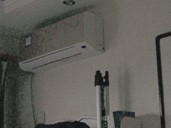} \\
        \texttt{Low-light level} $3$. \\
        \includegraphics[width=0.19\linewidth]{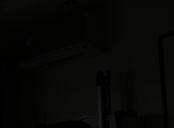} & 
        \includegraphics[width=0.19\linewidth]{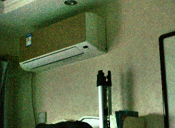} & 
        \includegraphics[width=0.19\linewidth]{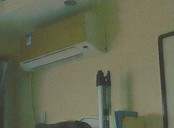} & 
        \includegraphics[width=0.19\linewidth]{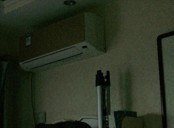} & 
        \includegraphics[width=0.19\linewidth]{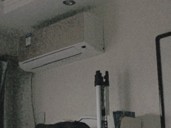} \\
        \texttt{Low-light level} $4$. \\
    \end{tabular}
    \caption{Visual comparison of low-light enhancement results for Scene 1: AC image. We compare with recent low-light enhancement methods: (a) low-light input image, (b) LIME, (c) deep-retinex, and (e)  Proposed. Notice that our method achieves better quality in the enhanced image than the other methods shown here.}
    \label{fig:visual_ac}
\end{figure*}

\begin{figure*}
    \centering
    \subfigure[LIME.]{\includegraphics[width=0.35\textwidth]{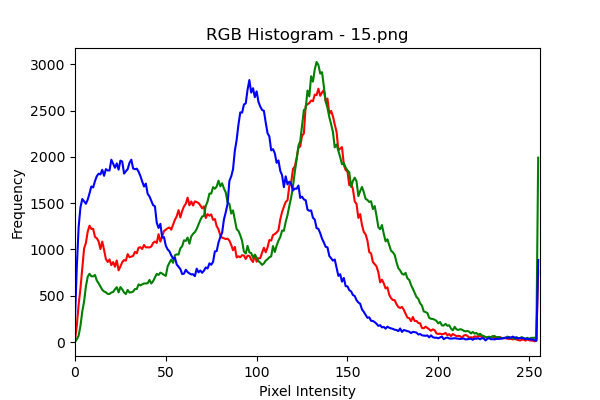}} 
    \subfigure[Deep-retinex.]{\includegraphics[width=0.35\textwidth]{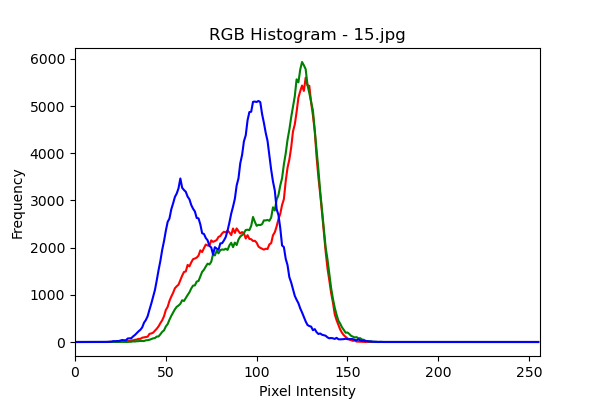}} \\
   \subfigure[Zero-DCE.]{\includegraphics[width=0.35\textwidth]{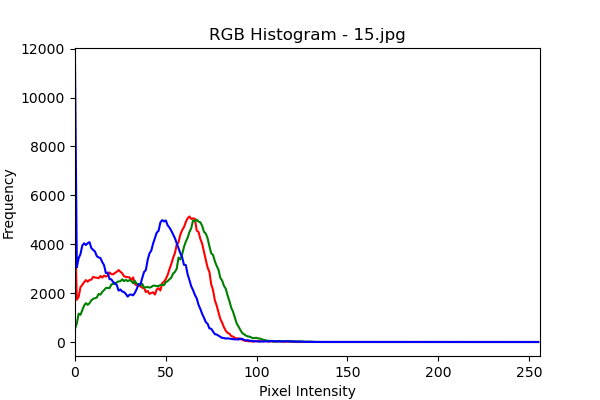}} 
   \subfigure[PIRL (proposed).]{\includegraphics[width=0.35\textwidth]{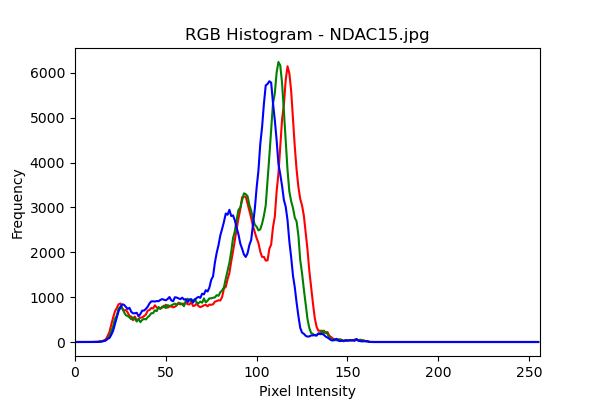}}
  \caption{Plot of histograms of RGB channels of recovered \textit{AC} image at low-light level 3 in Figure~\ref{fig:visual_ac}. An image with good color consistency exhibits very similar histograms for RGB colors. Notice that our method gives the best color consistency resulting no color distortion unlike other methods.}
    \label{fig:RGB_hist_ac}
\end{figure*}

\begin{figure*}
    \centering
    \setlength{\tabcolsep}{2pt}
    \renewcommand{\arraystretch}{1.2}
    \begin{tabular}{ccccc}
       (a)  \textbf{Input} & (b) \textbf{LIME} & (c) \textbf{Deep-Retinex} & (d) \textbf{Zero-DCE} & (e) \textbf{Proposed} \\
        \includegraphics[width=0.19\linewidth]{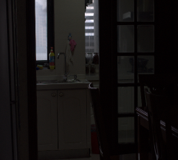} & 
        \includegraphics[width=0.19\linewidth]{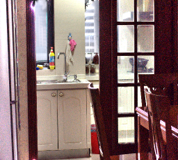} & 
        \includegraphics[width=0.19\linewidth]{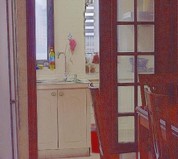} & 
        \includegraphics[width=0.19\linewidth]{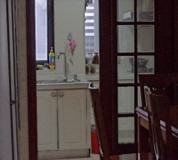} & 
        \includegraphics[width=0.19\linewidth]{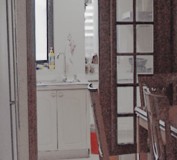} \\
        \texttt{Low-light level} $1$. \\
        \includegraphics[width=0.19\linewidth]{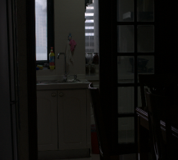} & 
        \includegraphics[width=0.19\linewidth]{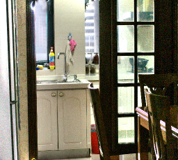} & 
        \includegraphics[width=0.19\linewidth]{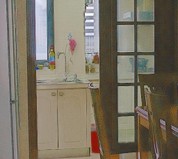} & 
        \includegraphics[width=0.19\linewidth]{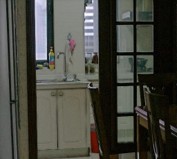} & 
        \includegraphics[width=0.19\linewidth]{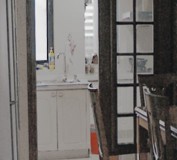} \\
        \texttt{Low-light level} $2$. \\
        \includegraphics[width=0.19\linewidth]{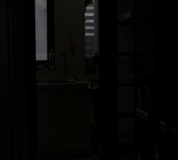} & 
        \includegraphics[width=0.19\linewidth]{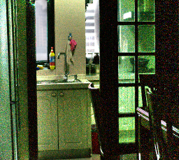} & 
        \includegraphics[width=0.19\linewidth]{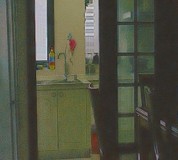} & 
        \includegraphics[width=0.19\linewidth]{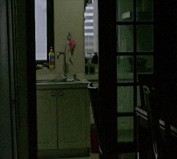} & 
        \includegraphics[width=0.19\linewidth]{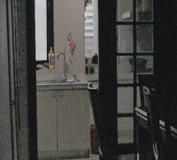} \\
        \texttt{Low-light level} $3$. \\
        \includegraphics[width=0.19\linewidth]{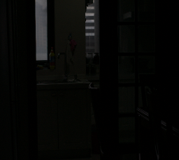} & 
        \includegraphics[width=0.19\linewidth]{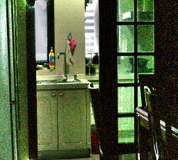} & 
        \includegraphics[width=0.19\linewidth]{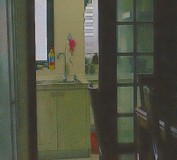} & 
        \includegraphics[width=0.19\linewidth]{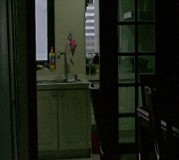} & 
        \includegraphics[width=0.19\linewidth]{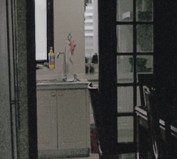} \\
        \texttt{Low-light level} $4$. \\
    \end{tabular}
    \caption{Visual comparison of low-light enhancement results on the \textit{Door} image:: (a) low-light input image, (b) LIME, (c) deep-retinex, (d) zero-DCE, and (e) Proposed method PIRL. Our method achieves significantly better visual quality in the enhanced image than the other methods shown.}
    \label{fig:visual2_door}
\end{figure*}

\subsubsection{Image Reconstruction (Output Layers)}

Our network produces three outputs through three separate branches. 
The pre-final output of the first branch (top output image shown in Figure \ref{fig:denoiseNet}) is passed through a sigmoid activation to constrain the values in the range $[0, 1]$. This output eventually generates a single-channel image that represents the illumination map. The second output is a three-channel image representing the reflectance map, also followed by a sigmoid activation. The third output represents the noise map consisting of three channels. In this case, we apply a \emph{tanh} activation function to allow negative and positive values to represent noise characteristics.

Notice that we estimate the Poisson noise (three-channel) separately under the assumption of extreme low-light imaging. It is important to note the novel retinex decomposition modeling in this work.

\subsection{Loss Function}
We propose a new loss function to efficiently train the proposed deep learning model. 
We optimize the training of our model by incorporating multiple components that guide the network performing image reconstruction, noise suppression, and smoothness regularization. Our loss function consists of multiple losses as follows:
%
\begin{equation}
    \mathcal{L}_{total} = \mathcal{L}_{rec} + \lambda_{1}\mathcal{L}_{ret} + \lambda_{2}\mathcal{L}_{n},
\end{equation}
 where \(\mathcal{L}_{rec}\), \(\mathcal{L}_{ret}\), \(\mathcal{L}_{n}\) denote reconstruction loss, Retinex estimation loss, and noise estimation loss respectively. The losses are weighted by hyper-parameters \(\lambda_{1}\), \(\lambda_{2}\).
 
 \subsubsection{Reconstruction Loss}
The error in decomposition is regulated by reconstruction loss function as follows:
\begin{equation}
    \mathcal{L}_{rec} = \parallel Y -(R\circ L\circ N)\parallel_{1}.
\end{equation}
Reconstruction loss measures the fidelity between the original and reconstructed images. This provides robustness to outliers and encourages sparsity in the reconstruction error.

\subsubsection{Retinex  Estimation Loss}
This loss component is expressed as a combination of loss due to decomposition loss $\mathcal{L}_{decom}$ and smoothness preserving smoothness loss \cite{chan2011augmented}  $\mathcal{L}_{sps}$ follows,
\begin{equation}
    \mathcal{L}_{ret} = \gamma \mathcal{L}_{decom}+(1-\gamma)\mathcal{L}_{sps}
\end{equation}
where \(\gamma\) is a hyperparameter that needed to be adjusted based on learning. 
In addition,
\begin{equation}
    \mathcal{L}_{decom} = \parallel X - R \circ L\parallel_{1}.
    \label{eq:Ldecom}
\end{equation}
Unlike the dark channel prior in the classical dehazing algorithm, or from the bright channel prior incorporated to impose constraints on the illumination map, our proposed deep network decomposes an image into $(R, L, N)$ without any prior for $R$ and $L$. However, we attempt to mitigate the noise component in the enhanced image by applying the loss $\mathcal{L}_{decom}$  in \eqref{eq:Ldecom}.

The smoothness preserving smoothness loss is given by:
\begin{equation}
    \mathcal{L}_{sps} = \left\|\Delta L \circ \ exp{(-\beta \Delta R)} \right\|_{1}
\end{equation}
whereas, \(\beta\) represents the coefficients for balanced structure and smoothness of illumination map, which is defined as \(10\) in our experiments and  \(\Delta R\) represents gradient of reflectance in both x and y direction, \(\Delta L\) represents gradient of Illumination in both x and y direction. The term exponential is employed to safeguard against structural loss in the illumination map, and gradient ensures the smoothness of the illumination.
An ideal illumination map should retain structural details from the input image while maintaining smoothness in texture. To achieve this, we introduce a structure-preserving smoothness loss \(L_{ret2}\), inspired by the Total Variation (TV) \cite{chan2011augmented}, which is used to weight the gradient of the reflectance map.

\subsubsection{Noise Loss}
The loss works as 
a constraint to extract the noisy component and mitigate noise in from the low light image;
\begin{equation}
    \mathcal{L}_{n} = \parallel N - \frac{\text{Poisson}(Y)}{Y + \alpha}\parallel_{1}
    \label{eq:PoissonLoss}
\end{equation}
which is used to suppress the intensity of noise (\(\alpha\) is set as \(10^{-6}\)  computational stability). Notice that the loss in \eqref{eq:PoissonLoss} is inspired by the existence of Poisson noise in extreme low-light images. 

We summarize the loss function in Figure~\ref{fig:loss}. Notice that introducing $\mathcal{L}_{n}$ loss is one of the novel contributions in our method. The hyperparameters $\gamma$, $\lambda_1$ and $\lambda_2$ are set using grid-search method during validation stage. 
Finally, we refer to our proposed method as \textbf{P}oisson \textbf{I}nformed \textbf{R}etinex of \textbf{L}ow-light images (PIRL).

\begin{table*}
\centering
\caption{Comparison for different methods on LOL dataset. The best are marked as \textcolor{red}{red}, the second best are marked as \textcolor{blue}{blue}.}
\begin{tabular}{|p{90pt}|p{90pt}|p{50pt}|p{50pt}|p{50pt}|}
\hline
Method & Reference & PSNR(dB)& SSIM & NIQE \\
\hline
LIME \cite{guo2016lime} & IEEE TIP 2016 & 15.66 & 0.42& 9.35  \\
LIME-filtered \cite{guo2016lime} & IEEE TIP 2016  &  16.77 & 0.59 & 4.53 \\
SRIE \cite{fu2016weighted}  & IEEE/CVF CVPR 2016  & 9.85 & 0.40& 7.72 \\
Deep Retinex \cite{Chen2018Retinex}  & BMVC 2018 &  16.06 & 0.58 &  4.31  \\
RRM \cite{li2018structure} & IEEE TIP 2018  & 13.88 & 0.658 & 5.810     \\
HybridNet \cite{ren2019low} & IEEE TIP 2019  & 16.60 & \textcolor{blue}{\bf{0.66}} & \textcolor{red}{\bf{3.42}} \\  
Zero-DCE \cite{guo2020zero} & IEEE/CVF CVPR 2020  & 12.35 & 0.55 & \textcolor{red}{\bf{3.42}}   \\
STAR \cite{xu2020star} & IEEE TIP 2020  & 12.64 & 0.538 & 6.205   \\
LR3M \cite{ren2020lr3m} & IEEE TIP 2020  & 10.22 & 0.399 & 7.522      \\
DRBN \cite{yang2020fidelity} & IEEE/CVF CVPR 2020  & 16.75 & 0.659 & 4.724      \\
RUAS \cite{liu2021retinex} & IEEE/CVF CVPR 2021  & 16.40 & 0.583 & 6.340      \\
RetinexDIP \cite{zhao2021retinexdip} & IEEE TCSVT 2021  & 9.442 & 0.334 & 7.070  \\
EFINet \cite{liu2022efinet} & IEEE TCSVT 2022  & 13.456 & 0.626 & - -     \\
SCL-LLE \cite{liang2022semantically}  & AAAI 2022  & 12.42 & 0.520 & 7.63     \\
SCI \cite{ma2022toward} & IEEE/CVF CVPR 2022  & 14.78 & 0.520 & 7.87     \\
SGZ \cite{zheng2022semantic} & IEEE/CVF WACV 2022  & 15.93 & 0.570 & 7.82     \\
UEGAN \cite{ni2020towards} & IEEE TIP 2020  & 12.22 & 0.262 & 5.627     \\
%
DUNP \cite{liang2022self} & IEEE TCSVT 2022  & 15.49 & 0.654 & 3.73     \\
GDP \cite{fei2023generative} & IEEE/CVF CVPR 2023  & 15.89 & 0.542 & 6.13     \\
NeurBR \cite{zhao2024non} & Pattern Recognition 2024  & 11.38 & 0.440 & 7.50     \\
PnP Retinex \cite{wu2024retinex}  & IEEE TNNLS 2024 & \textcolor{blue}{\bf{17.12}} & \textcolor{blue}{\bf{0.66}} & 3.95 \\
\hline
PNAR \cite{kong2021low} & IEEE SPL 2021  & 16.53 & 0.583 & - -      \\
IQA \cite{Huang2022QuaternionPoisson} & IEEE SPL 2022  & \textcolor{red}{\bf{17.95}} & 0.658 & - -    \\
\textbf{PIRL} & Proposed & \textcolor{blue}{\bf{17.12}}  & \textcolor{red}{\bf{0.67}} & \textcolor{blue}{\bf{3.54}}   \\
\hline
\end{tabular}
\label{tab:psnr_lol}
\end{table*}

\section{Experimental Results}
\label{sec:result}

\subsubsection{Dataset and Experimental Setup}
We used a publicly available Low Light (LOL) image dataset \cite{Chen2018Retinex} to train and test our deep learning method. The LOL dataset has 500 paired low-light and normal-light images. These images were captured by changing the exposure time and ISO, whereas the other configurations of the cameras are unchanged. 
The training process employs the model for the joint decomposition of low-light images into illumination, reflectance, and noise components.

During training, each input consists of \(3 \times 128\times 128\) sized RGB color patches of paired low-light and normal-light images.
The ADAM optimizer is utilized with parameter settings: learning rate ($lr$) initialized at $0.001$, stability factor $\epsilon = 10^{-8}$, and moving average control coefficients $\beta_{1}=0.9$ and $\beta_{2}=0.999$. The training is conducted for 100 epochs with a fixed batch size of 16. Then, we pre-processed the dataset by cropping the images into patches of $128 \times 128$. In addition, the  proposed algorithm is implemented on the PyTorch platform.

\subsubsection{Benchmarking Methods}
To demonstrate the effectiveness of our proposed method, we compare the results with state-of-the-art methods. We compare with both sub-types of classical and deep learning methods. 
The non-deep learning (i.e. classical) methods compared are: low-light image enhancement (LIME) \cite{guo2016lime}, LIME-filtered \cite{guo2016lime},  simultaneous reflectance and illumination estimation (SRIE) \cite{fu2016weighted}, robust retinex model (RRM) \cite{li2018structure}, structure and texture aware retinex (STAR) \cite{xu2020star},  and low-rank regularized retinex model (LR3M) \cite{ren2020lr3m}. The reference information in Table~\ref{tab:psnr_lol} suggests that most of the recent methods are in deep learning.

In this work, we compare our method with the following deep learning methods for low-light image enhancement: deep Retinex \cite{Chen2018Retinex}, zero-DCE \cite{guo2020zero}, hybridNet \cite{ren2019low}, deep recursive band network (DRBN) \cite{yang2020fidelity},  retinex-inspired unrolling with architecture search (RUAS) \cite{liu2021retinex}, RetinexDIP \cite{zhao2021retinexdip}, enhancement-fusion iterative network (EFINet) \cite{liu2022efinet},  semantically contrastive learning for low-light image enhancement (SCL-LLE) \cite{liang2022semantically}, self-calibrated illumination (SCI) \cite{ma2022toward}, semantic-guided zero-shot learning (SGZ) \cite{zheng2022semantic}, unsupervised
enhancement with generative adversarial network (UEGAN) \cite{ni2020towards}, discrepant untrained network priors (DUNP) \cite{liang2022self}, generative diffusion prior (GDP) \cite{fei2023generative}, NeurBR \cite{zhao2024non}. It is worth noting that we have included state-of-the-art artificial intelligence (AI) novel directions: generative adversarial network (GAN), diffusion method, zero-shot learning, etc. 

We also compared two recent methods \cite{kong2021low,Huang2022QuaternionPoisson} built on the idea of Poisson noise in low-light images. The authors in \cite{Huang2022QuaternionPoisson} introduced a regularizer based on the gamma-correction function and formulated it into a screened Poisson equation for the enhancement of low-light images. The iterative optimization algorithm in \cite{kong2021low} employed regularization terms for the piecewise smooth prior of the illumination component and the Poisson noise distribution prior, which contributes to noise suppression in different noise intensity. 

\subsubsection{Visual Comparison}
We show the visual results of our method along with the baseline in two different scenes with varying illumination levels.
In our experiments, we have explicitly considered input images with four different levels of illumination (severe low-lightness as we move down from the first row to the fourth row). 
In Figure \ref{fig:visual_ac}, we present the visual results of the enhancement of an \textit{AC} image that is poorly illuminated at four different levels of low-lightness. Our method produces the best restored visual quality images without color distortion as shown in Figure \ref{fig:visual_ac}(e). 
At first glance, we notice the evidence of color distortion by LIME and other methods shown in Figure~\ref{fig:visual_ac}. Moreover, the nature of color distortion is found to have a relation with the amount of low-lightness in the input image. Note that the color distortion gets worse with a very low level of illumination (as is the case for the last row), which is the case of extreme low-light conditions. Our method significantly outperforms the baseline methods in terms of producing color distortion-free images with high measure of naturalness. 

For a deeper understanding of the color distortion using various methods, we studied the histogram relationship of \{R, G, B\} channels. An image with good color consistency is expected to exhibit very similar histograms for RGB colors. In Figure~\ref{fig:RGB_hist_ac}, we show the \{R, G, B\} histograms of the restored images using LIME, deep-retinex, zero-DCE, and our method PIRL (for low-light level 3 in Figure~\ref{fig:visual_ac}).  Notice that our method gives the best color consistency resulting, no color distortion unlike other methods.

In Figure~\ref{fig:visual2_door}, we show the visual results of another low-light image suffered by four different levels of low-lightness. It is visually evident that the color distortion gets worse with a very low level of illumination, which is the case for extreme low-light conditions.  In both Figures \ref{fig:visual_ac} and \ref{fig:visual2_door}, we have used the same scene as input images but with varying exposure (along the rows). To be precise, as we move down from the first row to the fourth row, the exposure in the input image decreases.  A lesser exposure image, for example, in the fourth row, has huge color shift towards green channel of the image. This prominent color shift in state-of-the-art methods increases as the exposure decreases in the image. For the same level of exposure, there is  no color shift or distortion in the enhanced image produced by our method. 

Based on exhaustive experiments, we conclude that as more an image is captured in lower exposure, more it will suffer from color shift or distortion when illumination enhancement is done while considering denoising along with it as separate task. When treating enhancement task as denoising only and considering an underexposure image as Poisson noise, it leads to attaining image enhancement without having a prominent color shift.

%

\subsubsection{Quantitative Comparison}
We evaluated the performance  of various methods using three image quality metrics: peak signal-to-noise ratio (PSNR) \cite{mittal2012making}, structural similarity index measure (SSIM) \cite{wang2004image}, and no-reference image quality evaluator (NIQE) \cite{mittal2012making}. 
With the availability of ground-truth in the LOL dataset, we computed the PSNR and SSIM values between the ground-truth and the output images (of each method). For both PSNR and SSIM, a higher value  indicates better quality. For the NIQE metric, a lower value indicates better quality of the enhanced image. In Table \ref{tab:psnr_lol}, we summarize the performance of low-light enhancement  using our method and the baselines. Notice that our method PIRL achieves the second highest PSNR, (marginally) the highest SSIM, and the second best (lowest) NIQE values. We argue that overall our method achieves competitive performance with the baseline methods in terms of both quantitative measure and visual quality of the restored images. Experimental results confirm that our method is able to perform efficient low-light enhancement even under extreme low light conditions.

\section{Conclusion}
\label{sec:conc}
In this work, we introduced a new deep learning method to address the challenging task of extreme low-light image enhancement and denoising. We demonstrated with experimental finding that many baseline methods suffer from color distortion when the illumination level of the scene/object is extremely poor.  We introduced a novel image decomposition framework in contrast to Retinex model of image decomposition. Furthermore, our encoder-decoder-based deep learning method could produce superior recovery of extreme low-light images without any visible color distortion.  Given that Poisson noise dominates in extreme low-light images because of limited photon counts, specialized Poisson denoising techniques are essential for preserving image quality and structural details.

In future research, we will focus on developing denoising models that can robustly handle mixed noise distributions while enabling computationally efficient (light-weight network) solutions for real-time feasibility.
We also plan to integrate a Poisson denoising scheme directly on raw low-light images, which could offer a more lightweight and computationally efficient alternative.
Leveraging raw image enhancement techniques, we would be able to improve performance utilizing unprocessed sensor data, enabling more accurate noise modeling. This direction of research will offer better structural preservation and enhanced visual quality in detail, while maintaining real-time applicability.

\bibliographystyle{IEEEtran}
\bibliography{refs2}

\begin{thebibliography}{10}
\providecommand{\url}[1]{#1}
\csname url@samestyle\endcsname
\providecommand{\newblock}{\relax}
\providecommand{\bibinfo}[2]{#2}
\providecommand{\BIBentrySTDinterwordspacing}{\spaceskip=0pt\relax}
\providecommand{\BIBentryALTinterwordstretchfactor}{4}
\providecommand{\BIBentryALTinterwordspacing}{\spaceskip=\fontdimen2\font plus
\BIBentryALTinterwordstretchfactor\fontdimen3\font minus \fontdimen4\font\relax}
\providecommand{\BIBforeignlanguage}[2]{{%
\expandafter\ifx\csname l@#1\endcsname\relax
\typeout{** WARNING: IEEEtran.bst: No hyphenation pattern has been}%
\typeout{** loaded for the language `#1'. Using the pattern for}%
\typeout{** the default language instead.}%
\else
\language=\csname l@#1\endcsname
\fi
#2}}
\providecommand{\BIBdecl}{\relax}
\BIBdecl

\bibitem{grigorescu2020survey}
S.~Grigorescu, B.~Trasnea, T.~Cocias, and G.~Macesanu, ``A survey of deep learning techniques for autonomous driving,'' \emph{Journal of Field Robotics}, vol.~37, no.~3, pp. 362--386, 2020.

\bibitem{shrikhande2023face}
S.~Shrikhande, S.~Borse, and S.~Bhatlawande, ``Face recognition based attendance system,'' 2023.

\bibitem{chen2003minimum}
S.-D. Chen and A.~R. Ramli, ``Minimum mean brightness error bi-histogram equalization in contrast enhancement,'' \emph{IEEE Transactions on Consumer Electronics}, vol.~49, no.~4, pp. 1310--1319, 2003.

\bibitem{abdullah2007dynamic}
M.~Abdullah-Al-Wadud, M.~H. Kabir, M.~A.~A. Dewan, and O.~Chae, ``A dynamic histogram equalization for image contrast enhancement,'' \emph{IEEE Transactions on Consumer Electronics}, vol.~53, no.~2, pp. 593--600, 2007.

\bibitem{sen2010automatic}
D.~Sen and S.~K. Pal, ``Automatic exact histogram specification for contrast enhancement and visual system based quantitative evaluation,'' \emph{IEEE Transactions on Image Processing}, vol.~20, no.~5, pp. 1211--1220, 2010.

\bibitem{ibrahim2007brightness}
H.~Ibrahim and N.~S.~P. Kong, ``Brightness preserving dynamic histogram equalization for image contrast enhancement,'' \emph{IEEE Transactions on Consumer Electronics}, vol.~53, no.~4, pp. 1752--1758, 2007.

\bibitem{ma2023low}
Q.~Ma, Y.~Wang, and T.~Zeng, ``Low-light image enhancement via implicit priors regularized illumination optimization,'' \emph{IEEE Transactions on Computational Imaging}, vol.~9, pp. 944--953, 2023.

\bibitem{land1977retinex}
E.~H. Land, ``The retinex theory of color vision,'' \emph{Scientific American}, vol. 237, no.~6, pp. 108--129, 1977.

\bibitem{jobson1997properties}
D.~J. Jobson, Z.-u. Rahman, and G.~A. Woodell, ``Properties and performance of a center/surround retinex,'' \emph{IEEE Transactions on Image Processing}, vol.~6, no.~3, pp. 451--462, 1997.

\bibitem{jobson1997multiscale}
------, ``A multiscale retinex for bridging the gap between color images and the human observation of scenes,'' \emph{IEEE Transactions on Image processing}, vol.~6, no.~7, pp. 965--976, 1997.

\bibitem{elad2005retinex}
M.~Elad, ``Retinex by two bilateral filters,'' \emph{Proc. International Conference on Scale-space Theories in Computer Vision}, pp. 217--229, 2005.

\bibitem{ghosh2019fast}
S.~Ghosh and K.~N. Chaudhury, ``Fast bright-pass bilateral filtering for low-light enhancement,'' \emph{Proc. IEEE International Conference on Image Processing (ICIP)}, pp. 205--209, 2019.

\bibitem{guo2016lime}
X.~Guo, Y.~Li, and H.~Ling, ``{LIME}: Low-light image enhancement via illumination map estimation,'' \emph{IEEE Transactions on Image Processing}, vol.~26, no.~2, pp. 982--993, 2016.

\bibitem{zhang2023learning}
Y.~Zhang, B.~Teng, D.~Yang, Z.~Chen, H.~Ma, G.~Li, and W.~Ding, ``Learning a single convolutional layer model for low light image enhancement,'' \emph{IEEE Transactions on Circuits and Systems for Video Technology}, vol.~34, no.~7, pp. 5995--6008, 2023.

\bibitem{shi2022unsupervised}
Y.~Shi, B.~Wang, X.~Wu, and M.~Zhu, ``Unsupervised low-light image enhancement by extracting structural similarity and color consistency,'' \emph{IEEE Signal Processing Letters}, vol.~29, pp. 997--1001, 2022.

\bibitem{xu2022structure}
K.~Xu, H.~Chen, C.~Xu, Y.~Jin, and C.~Zhu, ``Structure-texture aware network for low-light image enhancement,'' \emph{IEEE Transactions on Circuits and Systems for Video Technology}, vol.~32, no.~8, pp. 4983--4996, 2022.

\bibitem{huang2022towards}
H.~Huang, W.~Yang, Y.~Hu, J.~Liu, and L.-Y. Duan, ``Towards low light enhancement with raw images,'' \emph{IEEE Transactions on Image Processing}, vol.~31, pp. 1391--1405, 2022.

\bibitem{li2023pixel}
X.~Li, M.~Liu, and Q.~Ling, ``Pixel-wise gamma correction mapping for low-light image enhancement,'' \emph{IEEE Transactions on Circuits and Systems for Video Technology}, vol.~34, no.~2, pp. 681--694, 2023.

\bibitem{wang2023enlighten}
H.~Wang, C.~Shu, and X.~Li, ``Enlighten fusion multiscale network for infrared and visible image fusion in dark environments,'' \emph{IEEE Signal Processing Letters}, vol.~30, pp. 1167--1171, 2023.

\bibitem{wang2023multi}
X.~Wang, Y.~Lin, and S.~Zhang, ``Multi-stream progressive restoration for low-light light field enhancement and denoising,'' \emph{IEEE Transactions on Computational Imaging}, vol.~9, pp. 70--82, 2023.

\bibitem{xie2024residual}
C.~Xie, L.~Fei, H.~Tao, Y.~Hu, W.~Zhou, J.~T. Hoe, W.~Hu, and Y.-P. Tan, ``Residual quotient learning for zero-reference low-light image enhancement,'' \emph{IEEE Transactions on Image Processing}, 2024.

\bibitem{wang2024extracting}
H.~Wang, X.~Yan, X.~Hou, K.~Zhang, and Y.~Dun, ``Extracting noise and darkness: Low-light image enhancement via dual prior guidance,'' \emph{IEEE Transactions on Circuits and Systems for Video Technology}, 2024.

\bibitem{he2025zero}
J.~He, S.~Palaiahnakote, A.~Ning, and M.~Xue, ``Zero-shot low-light image enhancement via joint frequency domain priors guided diffusion,'' \emph{IEEE Signal Processing Letters}, 2025.

\bibitem{Chen2018Retinex}
W.~Y. J.~L. Chen~Wei, Wenjing~Wang, ``Deep retinex decomposition for low-light enhancement,'' \emph{Proc. British Machine Vision Conference}, 2018.

\bibitem{danielyan2011bm3d}
A.~Danielyan, V.~Katkovnik, and K.~Egiazarian, ``{BM3D} frames and variational image deblurring,'' \emph{IEEE Transactions on Image Processing}, vol.~21, no.~4, pp. 1715--1728, 2011.

\bibitem{ko2021learning}
S.~Ko, J.~Park, B.~Chae, and D.~Cho, ``Learning lightweight low-light enhancement network using pseudo well-exposed images,'' \emph{IEEE Signal Processing Letters}, vol.~29, pp. 289--293, 2021.

\bibitem{yang2022rethinking}
S.~Yang, D.~Zhou, J.~Cao, and Y.~Guo, ``Rethinking low-light enhancement via transformer-{GAN},'' \emph{IEEE Signal Processing Letters}, vol.~29, pp. 1082--1086, 2022.

\bibitem{xu2022rawformer}
W.~Xu, X.~Dong, L.~Ma, A.~B.~J. Teoh, and Z.~Lin, ``Rawformer: An efficient vision transformer for low-light raw image enhancement,'' \emph{IEEE Signal Processing Letters}, vol.~29, pp. 2677--2681, 2022.

\bibitem{shen2022blind}
L.~Shen, Z.~Ma, M.~J. Er, Y.~Fan, and Q.~Yin, ``Blind adaptive structure-preserving imaging enhancement for low-light condition,'' \emph{IEEE Signal Processing Letters}, vol.~29, pp. 917--921, 2022.

\bibitem{Guan2019NODEEL}
H.~Guan, L.~Liu, S.~J. Moran, F.~Song, and G.~G. Slabaugh, ``Node: Extreme low light raw image denoising using a noise decomposition network,'' \emph{ArXiv}, vol. abs/1909.05249, 2019.

\bibitem{hasinoff2021photon}
S.~W. Hasinoff, ``Photon, {P}oisson noise,'' in \emph{Computer vision: a reference guide}.\hskip 1em plus 0.5em minus 0.4em\relax Springer, 2021, pp. 980--982.

\bibitem{gnanasambandam2024secrets}
A.~Gnanasambandam, Y.~Sanghvi, and S.~H. Chan, ``The secrets of non-blind {P}oisson deconvolution,'' \emph{IEEE Transactions on Computational Imaging}, 2024.

\bibitem{foi2008practical}
A.~Foi, M.~Trimeche, V.~Katkovnik, and K.~Egiazarian, ``Practical {Poissonian-Gaussian} noise modeling and fitting for single-image raw-data,'' \emph{IEEE Transactions on Image Processing}, vol.~17, no.~10, pp. 1737--1754, 2008.

\bibitem{niknejad2018poisson}
M.~Niknejad and M.~A. Figueiredo, ``Poisson image denoising using best linear prediction: a post-processing framework,'' \emph{Proc. 26th European Signal Processing Conference (EUSIPCO)}, pp. 2230--2234, 2018.

\bibitem{wang2017fast}
W.~Wang and C.~He, ``A fast and effective algorithm for a {P}oisson denoising model with total variation,'' \emph{IEEE Signal Processing Letters}, vol.~24, no.~3, pp. 269--273, 2017.

\bibitem{talbot2009efficient}
H.~Talbot, H.~Phelippeau, M.~Akil, and S.~Bara, ``Efficient {P}oisson denoising for photography,'' \emph{Proc. 16th IEEE International Conference on Image Processing (ICIP)}, pp. 3881--3884, 2009.

\bibitem{kong2021low}
X.-Y. Kong, L.~Liu, and Y.-S. Qian, ``Low-light image enhancement via {Poisson} noise aware retinex model,'' \emph{IEEE Signal Processing Letters}, vol.~28, pp. 1540--1544, 2021.

\bibitem{huang2022quaternion}
C.~Huang, Y.~Fang, T.~Wu, T.~Zeng, and Y.~Zeng, ``Quaternion screened {Poisson} equation for low-light image enhancement,'' \emph{IEEE Signal Processing Letters}, vol.~29, pp. 1417--1421, 2022.

\bibitem{zhao2025deep}
Q.~Zhao, G.~Li, B.~He, and R.~Shen, ``Deep learning for low-light vision: A comprehensive survey,'' \emph{IEEE Transactions on Neural Networks and Learning Systems}, 2025.

\bibitem{liu2024low}
X.~Liu, Q.~Xie, Q.~Zhao, H.~Wang, and D.~Meng, ``Low-light image enhancement by retinex-based algorithm unrolling and adjustment,'' \emph{IEEE Transactions on Neural Networks and Learning Systems}, vol.~35, no.~11, pp. 15\,758--15\,771, 2024.

\bibitem{zhao2024non}
Z.~Zhao, H.~Lin, D.~Shi, and G.~Zhou, ``A non-regularization self-supervised retinex approach to low-light image enhancement with parameterized illumination estimation,'' \emph{Pattern Recognition}, vol. 146, p. 110025, 2024.

\bibitem{fei2023generative}
B.~Fei, Z.~Lyu, L.~Pan, J.~Zhang, W.~Yang, T.~Luo, B.~Zhang, and B.~Dai, ``Generative diffusion prior for unified image restoration and enhancement,'' in \emph{Proc. IEEE/CVF Conference on Computer Vision and Pattern Recognition}, 2023, pp. 9935--9946.

\bibitem{liang2022self}
J.~Liang, Y.~Xu, Y.~Quan, B.~Shi, and H.~Ji, ``Self-supervised low-light image enhancement using discrepant untrained network priors,'' \emph{IEEE Transactions on Circuits and Systems for Video Technology}, vol.~32, no.~11, pp. 7332--7345, 2022.

\bibitem{guo2020zero}
C.~Guo, C.~Li, J.~Guo, C.~C. Loy, J.~Hou, S.~Kwong, and R.~Cong, ``Zero-reference deep curve estimation for low-light image enhancement,'' \emph{Proc. IEEE/CVF Conference on Computer Vision and Pattern Recognition}, pp. 1780--1789, 2020.

\bibitem{9334429}
Y.~Jiang, X.~Gong, D.~Liu, Y.~Cheng, C.~Fang, X.~Shen, J.~Yang, P.~Zhou, and Z.~Wang, ``{EnlightenGAN}: Deep light enhancement without paired supervision,'' \emph{IEEE Transactions on Image Processing}, vol.~30, pp. 2340--2349, 2021.

\bibitem{Zhang2023SingleLayer}
Y.~Zhang, B.~Teng, D.~Yang, Z.~Chen, H.~Ma, G.~Li, and W.~Ding, ``Learning a single convolutional layer model for low light image enhancement,'' \emph{IEEE Transactions on Circuits and Systems for Video Technology}, vol.~34, no.~7, pp. 5995--6008, 2023.

\bibitem{Kong2021PoissonRetinex}
X.-Y. Kong, L.~Liu, and Y.-S. Qian, ``Low-light image enhancement via {Poisson} noise aware retinex model,'' \emph{IEEE Signal Processing Letters}, vol.~28, pp. 1540--1544, 2021.

\bibitem{Huang2022QuaternionPoisson}
C.~Huang, Y.~Fang, T.~Wu, T.~Zeng, and Y.~Zeng, ``Quaternion screened poisson equation for low-light image enhancement,'' \emph{IEEE Signal Processing Letters}, vol.~29, pp. 1417--1421, 2022.

\bibitem{hines2008probability}
W.~W. Hines, D.~C. Montgomery, and D.~M. G. C.~M. Borror, \emph{{P}robability and Statistics in Engineering}.\hskip 1em plus 0.5em minus 0.4em\relax John Wiley \& Sons, 2008.

\bibitem{lukac2005color}
R.~Lukac, K.~N. Plataniotis, and D.~Hatzinakos, ``Color image zooming on the {Bayer} pattern,'' \emph{IEEE Transactions on Circuits and Systems for Video Technology}, vol.~15, no.~11, pp. 1475--1492, 2005.

\bibitem{sonka2013image}
M.~Sonka, V.~Hlavac, and R.~Boyle, \emph{Image Processing, Analysis and Machine Vision}.\hskip 1em plus 0.5em minus 0.4em\relax Springer, 2013.

\bibitem{ma2024low}
S.~Ma, W.~Pan, N.~Li, S.~Du, H.~Liu, B.~Xu, C.~Xu, and X.~Li, ``Low-light image enhancement using retinex-based network with attention mechanism.'' \emph{International Journal of Advanced Computer Science \& Applications}, vol.~15, no.~1, 2024.

\bibitem{parabextreme}
N.~Parab, ``Extreme low-light single-image denoising model.''

\bibitem{8588010}
T.~Piriyatharawet, W.~Kumwilaisak, and P.~Lasang, ``Image denoising with deep convolutional and multi-directional {LSTM} networks under {P}oisson noise environments,'' \emph{Proc. 18th International Symposium on Communications and Information Technologies (ISCIT)}, pp. 90--95, 2018.

\bibitem{chan2011augmented}
S.~H. Chan, R.~Khoshabeh, K.~B. Gibson, P.~E. Gill, and T.~Q. Nguyen, ``An augmented {Lagrangian} method for total variation video restoration,'' \emph{IEEE Transactions on Image Processing}, vol.~20, no.~11, pp. 3097--3111, 2011.

\bibitem{fu2016weighted}
X.~Fu, D.~Zeng, Y.~Huang, X.-P. Zhang, and X.~Ding, ``A weighted variational model for simultaneous reflectance and illumination estimation,'' \emph{Proc. IEEE Conference on Computer Vision and Pattern Recognition}, pp. 2782--2790, 2016.

\bibitem{li2018structure}
M.~Li, J.~Liu, W.~Yang, X.~Sun, and Z.~Guo, ``Structure-revealing low-light image enhancement via robust retinex model,'' \emph{IEEE Transactions on Image Processing}, vol.~27, no.~6, pp. 2828--2841, 2018.

\bibitem{ren2019low}
W.~Ren, S.~Liu, L.~Ma, Q.~Xu, X.~Xu, X.~Cao, J.~Du, and M.-H. Yang, ``Low-light image enhancement via a deep hybrid network,'' \emph{IEEE Transactions on Image Processing}, vol.~28, no.~9, pp. 4364--4375, 2019.

\bibitem{xu2020star}
J.~Xu, Y.~Hou, D.~Ren, L.~Liu, F.~Zhu, M.~Yu, H.~Wang, and L.~Shao, ``Star: A structure and texture aware retinex model,'' \emph{IEEE Transactions on Image Processing}, vol.~29, pp. 5022--5037, 2020.

\bibitem{ren2020lr3m}
X.~Ren, W.~Yang, W.-H. Cheng, and J.~Liu, ``{LR3M}: Robust low-light enhancement via low-rank regularized retinex model,'' \emph{IEEE Transactions on Image Processing}, vol.~29, pp. 5862--5876, 2020.

\bibitem{yang2020fidelity}
W.~Yang, S.~Wang, Y.~Fang, Y.~Wang, and J.~Liu, ``From fidelity to perceptual quality: A semi-supervised approach for low-light image enhancement,'' in \emph{Proc. IEEE/CVF Conference on Computer Vision and Pattern Recognition}, 2020, pp. 3063--3072.

\bibitem{liu2021retinex}
R.~Liu, L.~Ma, J.~Zhang, X.~Fan, and Z.~Luo, ``Retinex-inspired unrolling with cooperative prior architecture search for low-light image enhancement,'' in \emph{Proc. IEEE/CVF Conference on Computer Vision and Pattern Recognition}, 2021, pp. 10\,561--10\,570.

\bibitem{zhao2021retinexdip}
Z.~Zhao, B.~Xiong, L.~Wang, Q.~Ou, L.~Yu, and F.~Kuang, ``{RetinexDIP}: A unified deep framework for low-light image enhancement,'' \emph{IEEE Transactions on Circuits and Systems for Video Technology}, vol.~32, no.~3, pp. 1076--1088, 2021.

\bibitem{liu2022efinet}
C.~Liu, F.~Wu, and X.~Wang, ``Efinet: Restoration for low-light images via enhancement-fusion iterative network,'' \emph{IEEE Transactions on Circuits and Systems for Video Technology}, vol.~32, no.~12, pp. 8486--8499, 2022.

\bibitem{liang2022semantically}
D.~Liang, L.~Li, M.~Wei, S.~Yang, L.~Zhang, W.~Yang, Y.~Du, and H.~Zhou, ``Semantically contrastive learning for low-light image enhancement,'' in \emph{Proc. AAAI conference on Artificial Intelligence}, vol.~36, no.~2, 2022, pp. 1555--1563.

\bibitem{ma2022toward}
L.~Ma, T.~Ma, R.~Liu, X.~Fan, and Z.~Luo, ``Toward fast, flexible, and robust low-light image enhancement,'' in \emph{Proc. IEEE/CVF Conference on Computer Vision and Pattern Recognition}, 2022, pp. 5637--5646.

\bibitem{zheng2022semantic}
S.~Zheng and G.~Gupta, ``Semantic-guided zero-shot learning for low-light image/video enhancement,'' in \emph{Proc. IEEE/CVF Winter Conference on Applications of Computer Vision}, 2022, pp. 581--590.

\bibitem{ni2020towards}
Z.~Ni, W.~Yang, S.~Wang, L.~Ma, and S.~Kwong, ``Towards unsupervised deep image enhancement with generative adversarial network,'' \emph{IEEE Transactions on Image Processing}, vol.~29, pp. 9140--9151, 2020.

\bibitem{wu2024retinex}
T.~Wu, W.~Wu, Y.~Yang, F.-L. Fan, and T.~Zeng, ``Retinex image enhancement based on sequential decomposition with a plug-and-play framework,'' \emph{IEEE Transactions on Neural Networks and Learning Systems}, vol.~35, no.~10, pp. 14\,559--14\,572, 2024.

\bibitem{mittal2012making}
A.~Mittal, R.~Soundararajan, and A.~C. Bovik, ``Making a “completely blind” image quality analyzer,'' \emph{IEEE Signal Processing Letters}, vol.~20, no.~3, pp. 209--212, 2012.

\bibitem{wang2004image}
Z.~Wang, A.~C. Bovik, H.~R. Sheikh, and E.~P. Simoncelli, ``Image quality assessment: from error visibility to structural similarity,'' \emph{IEEE Transactions on Image Processing}, vol.~13, no.~4, pp. 600--612, 2004.

\end{thebibliography}

\end{document}